\documentclass{appolb}
\usepackage{psfig}

\begin{document}
\title{Omega photoproduction.
\thanks{Presented at Meson 2002, 7th Int. Workshop on Production, 
Properties and Interaction of Mesons, Cracow, Poland, May 24-28 2002.}
}
\author{A. Sibirtsev, S. Krewald
\address{Institut f\"ur Kernphysik, Forschungszentrum J\"ulich}
\and
K. Tsushima
\address{ Department of Physics and Astronomy, University
of Georgia, Athens, USA}
}
\maketitle
\begin{abstract}
Photoproduction of $\omega$ is analyzed within meson exchange model
and Regge model and compared to $\rho$ photoproduction. 
An interplay between two models and uncertainties in data
reproduction are discussed. 
\end{abstract}

The vector meson photoproduction at small momentum transfers $|t|$ or
$|u|$  is traditionally discussed in terms of the Regge model.  
Recent CLAS data on $\rho$ photoproduction~\cite{Battaglieri} at 
low $|t|$ indicate that at low energies the dominant contribution 
comes from  $f_2$ exchanges, while at high 
energies~\cite{Donnachie1} it is due to Pomeron exchange. 
ZEUS data~\cite{Breitweg} on $\rho$ photoproduction 
require an additional contribution from  hard 
Pomeron exchange~\cite{Donnachie1}.
The $\phi$ photoproduction~\cite{Anciant} even at low energies
at small $|t|$ is dominated  by Pomeron exchange because of the 
$s{\bar s}$ structure of the $\phi$-meson.
At backward angles, where  $|u|$ is small, the $\rho$-meson
photoproduction is given by the exchange of nucleon and $\Delta$ Regge 
trajectories in the $u$ channel. The backward $\phi$-meson
photoproduction is due to the $u$-channel nucleon exchange.

While $\rho$ and $\phi$ photoproduction were 
studied systematically, the $\omega$ photoproduction has
been analyzed very selectively. It was found~\cite{Laget1,Sibirtsev2} that
Regge model calculation by $\simeq$26\% underestimate experimental 
data on total $\gamma{+}p{\to}\omega {+}p$ cross section 
at $E_\gamma{\ge}$20~GeV. This discrepancy is disturbing since
it is strongly believed that at high energies the
Pomeron exchange should be able to describe 
the  $\omega$ photoproduction as well as  available  
data on $\rho$ and $\phi$ photoproduction. 

\begin{figure}[t]
\hspace*{-3mm}\psfig{file=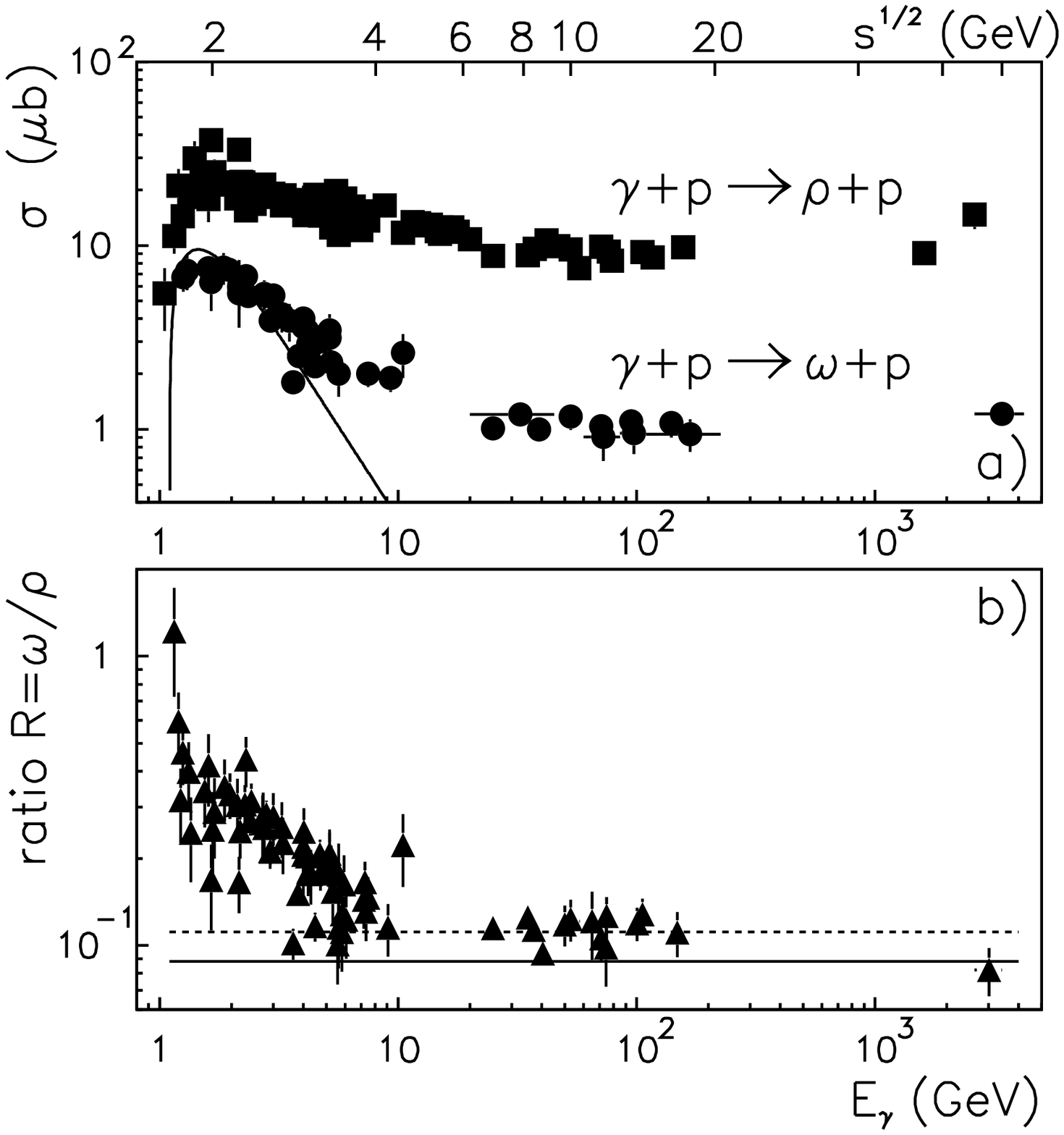,width=7cm,height=6.4cm}
\hspace*{-8mm}\psfig{file=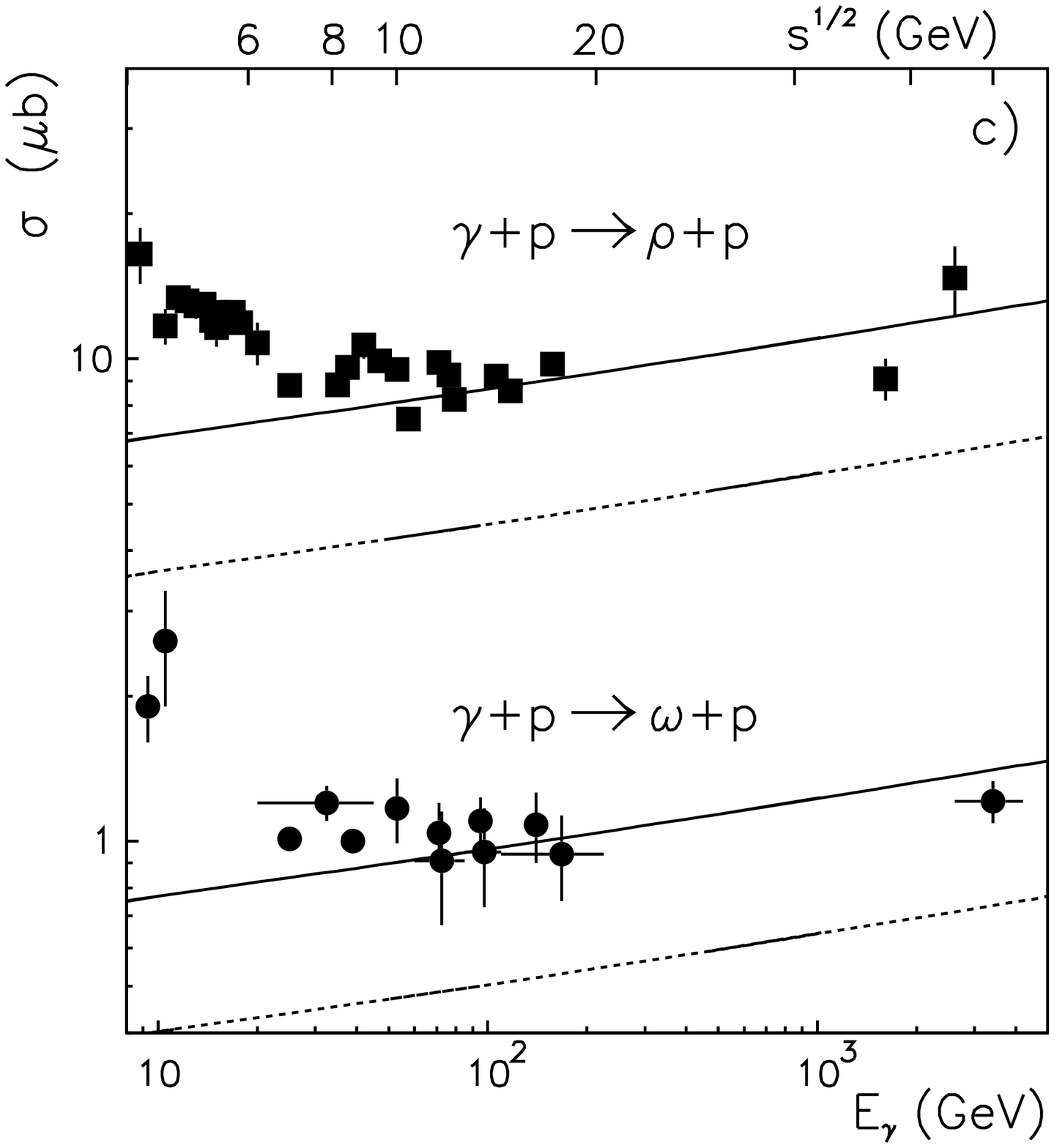,width=7cm,height=6.6cm}
\vspace*{-8mm}
\caption[]{Figures a) and b) show total $\gamma{+}p{\to}\rho{+}p$ and
$\gamma{+}p{\to}\omega{+}p$ cross section as
a function of  photon energy $E_\gamma$ (lower axis) or
invariant collision energy $\sqrt{s}$ (upper axis). The
solid line in a) shows the meson exchange calculations.
The solid lines in c) indicate the Pomeron exchange calculations 
with cut-off parameter $\mu_0$=1.1~GeV and 
coupling $\beta_0$=2.35~GeV$^{-1}$, while the dashed lines 
are the results for $\beta_0$=2.0~GeV$^{-1}$. Figure b) shows the
ratio of the total $\omega$ and $\rho$ photoproduction cross sections,
the solid line is the ratio of $\omega\gamma$ to $\rho\gamma$   couplings
determined by their $e^+e^-$ decay widths, while the dashed line indicates 
the expected SU(3) ratio.}
\label{reg12}
\end{figure}
The Pomeron exchange amplitude ${\cal T}_P$ 
for vector meson photoproduction is explicitly
given~\cite{Donnachie1,Laget1} as
\begin{equation}
{\cal T}_{P}{=}3F_1(t)\, \frac{8 \sqrt{6}\, m_q e_q f_{\gamma{V}} \beta_0^2}
{4m_q^2-t}\,  (\varepsilon{\cdot}\varepsilon_V) 
s \left( \frac{s}{s_0}
\right)^{\alpha_{P}(t)-1} \!\!\!\!\!
\frac{\mu_0^2}{2\mu_0^2+4m_q^2-t},
\label{pom}
\end{equation}
where $F_1(t)$ is  proton isoscalar EM form factor,
$e_q$ and $m_q$ are the quark charge and mass, while
$\varepsilon$ and $\varepsilon_V$ are the polarization vectors of 
the photon and vector meson, respectively and
$f_{\gamma{V}}$ is the coupling constant 
given by $V{\to}l^+l^-$ decay width. Furthermore, $s$ is the squared invariant
collision energy, $\alpha_P{=}1.08+0.25t$ is the Pomeron trajectory,
$s_0$=4~GeV$^2$ and $\beta_0$ determines 
the strength of the effective Pomeron coupling to the quark,
while $\mu_0$ accounts that the coupling to off-shell
quark is not pointlike but dressed with the form factor given 
by the last term of Eq.(\ref{pom}). 

It is clear that within the Pomeron exchange model\footnote{As well as for 
all models accounting for the ${\gamma\omega}$ and ${\gamma\rho}$ couplings
given by their experimental dileptonic  decay widths.}
the ratio of  $\gamma{+}p\to\omega{+}p$
to $\gamma{+}p{\to}\rho{+}p$ cross sections is driven  by the
ratio of $f^2_{\gamma\omega}{/}f^2_{\gamma\rho}$ coupling constants
squared, which are determined by relevant dileptonic decay widths. 
The experimental ratio
given by $V{\to}e^+e^-$ decay  is  0.088$\pm$0.005 and it is 
different from the SU(3) ratio given by 1/9.

Fig.\ref{reg12}a) shows the total $\rho$ and $\omega$ photoproduction
cross sections, while Fig.\ref{reg12}b) illustrates the
ratio $R$ of the total $\gamma{+}p{\to}\omega{+}p$
to $\gamma{+}p{\to}\rho{+}p$ cross sections. Obviously at 
$E_\gamma{\ge}$6~GeV experimental data are consistent with $R$=1/9, but they
are underestimated by 26\% as compared with ratio 
given by experimental $V{\to}e^+e^-$ decay width. This discrepancy
can not be addressed to the finite $\rho$  width 
correction~\cite{Renard} (factor $\simeq$1.1) or to the standard $\rho$
vector dominance model correction~\cite{Weise} ($\simeq$1.25)
and still remains an open problem. One of the possible explanation
is the $\rho{-}\omega$ mixing or $\rho{N}{\to}\omega{N}$ transition 
due to final state interaction. In that case the 
photoproduced $\rho$-mesons can be converted to detected 
$\omega$-mesons, which might account for an additional 
$\simeq$26\% for $\omega$ production rate. 

\begin{figure}[t]
\hspace*{-4mm}\psfig{file=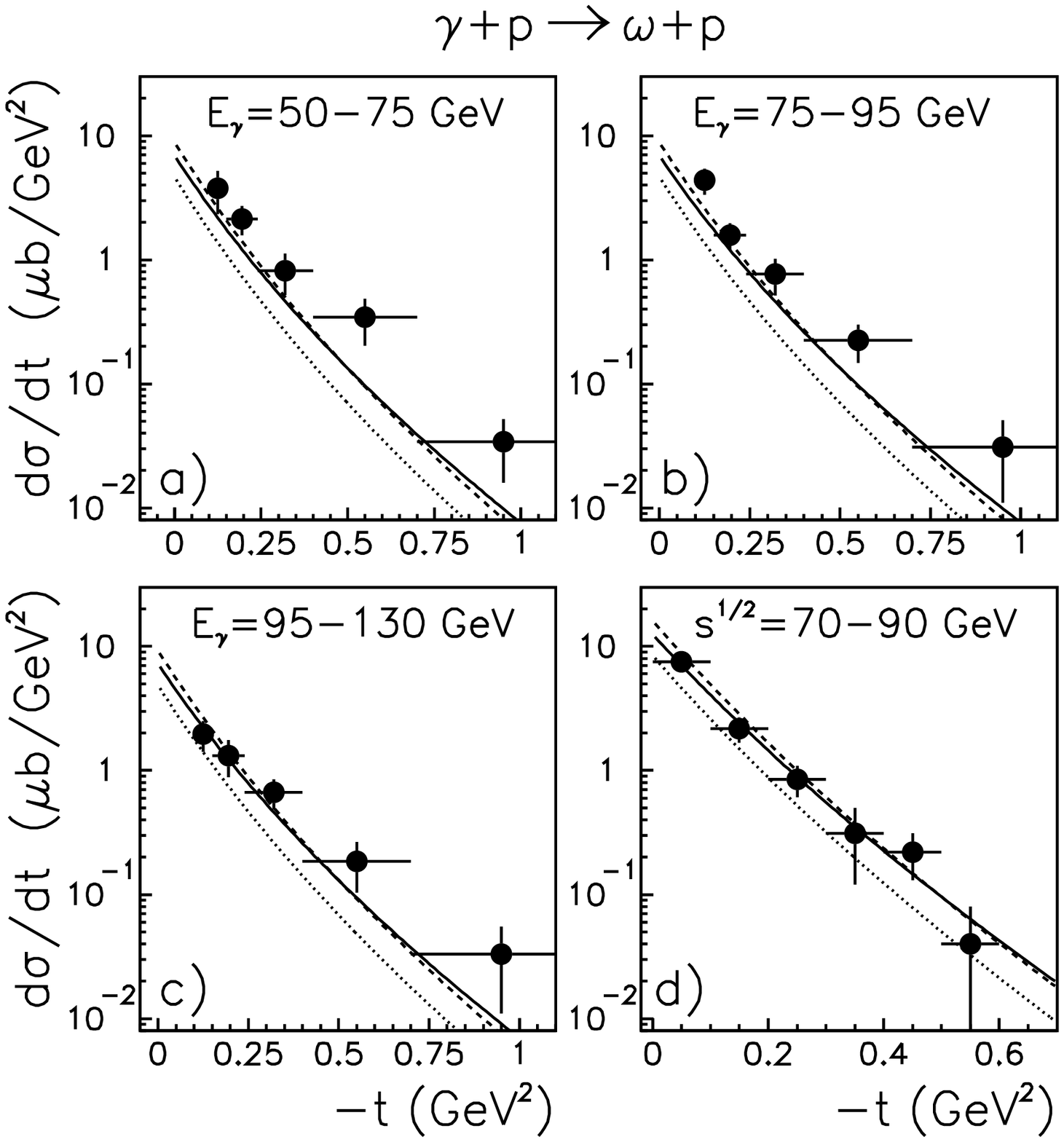,width=8.cm,height=7.1cm}
\hspace*{-10mm}\psfig{file=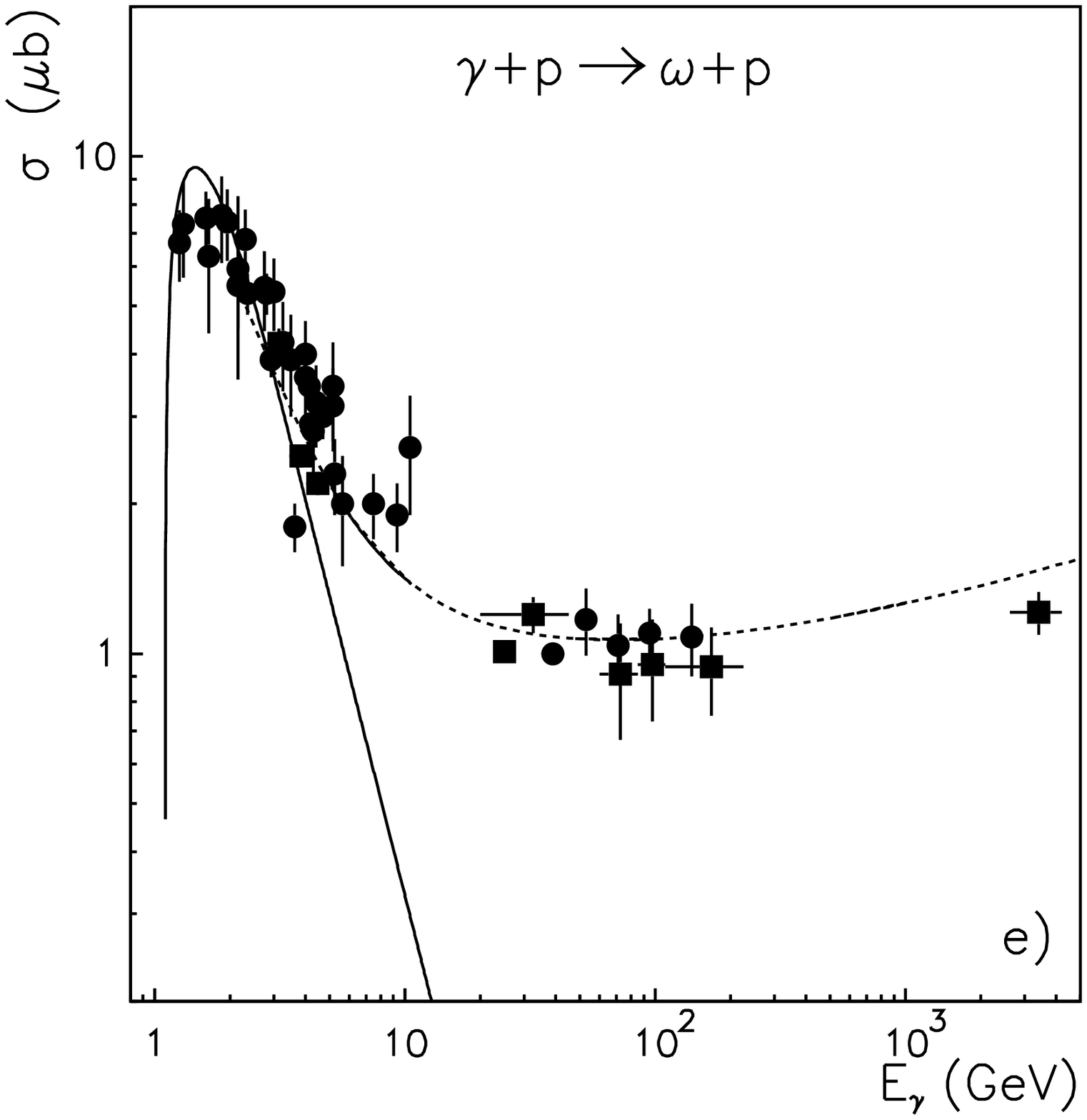,width=5.9cm,height=7.1cm}
\vspace*{-8mm}
\caption[]{Figures a)-d) show differential $\gamma{+}p{\to}\omega{+}p$ 
cross section as a function of four momentum transfer squared  
at different photon energies. The lines show the calculations by Pomeron 
exchange with parameters: $\beta_0$=2.0~GeV$^{-1}$, 
$\mu_0$=1.1~GeV -dotted, $\beta_0$=2.35~GeV$^{-1}$, $\mu_0$=1.1~GeV 
-dashed  and $\beta_0$=2.0~GeV$^{-1}$, $\mu_0$=2.5~GeV -solid. Figure
e) shows the energy dependence of the total $\omega$ photoproduction cross 
section. Solid line is the meson exchange calculation, while the dashed line
shows the Regge model result obtained with $\pi$, $f_2$ and 
soft and hard Pomeron exchanges.}
\label{reg5a}
\end{figure}

Free parameters of Pomeron exchange are the Pomeron-quark 
coupling $\beta_0$ and form factor cut-off $\mu_0$. The
coupling $\beta_0$=2.0~GeV$^{-1}$ was deduced~\cite{Donnachie1} from 
$pp$ scattering, while $\beta_0$=2.35~GeV$^{-1}$
was obtained~\cite{Pichowsky} from an analysis of $\pi{p}$ 
elastic scattering at high energies. 
The cut-off  $\mu_0$ can be fitted to photoproduction data.
The calculations by the Pomeron exchange model with experimental values
$\beta_0$=2.0~GeV$^{-1}$ and $\mu_0$=1.1~GeV~\cite{Donnachie1}
are shown by the dashed lines in Fig.\ref{reg12}c) and substantially
underestimate experimental data both for $\rho$ and $\omega$
photoproduction. Here we used the $\gamma\omega$ and $\gamma\rho$
couplings from SU(3). The solid line show our results obtained with
the coupling constant $\beta_0$=2.35~GeV$^{-1}$. 
Data can be as well fixed by adjusting the cut-off
parameter $\mu_0$=2.5~GeV, as is illustrated by Fig.\ref{reg5a}a)-d).
Here the data on differential $\gamma{p}{\to}{\omega}p$
cross section are compared to calculations.
Obviously, the data can be well reproduced by varying both
$\beta_0$ and $\mu_0$ and for this reason it is impossible
to fix the Pomeron-quark coupling by photoproduction data
alone.

Moreover,  Fig.\ref{reg12}b) indicates strong differences between 
$\rho$ and $\omega$ photoproduction at $E_\gamma{\le}$6~GeV.
Substantial enhancement of $\omega$ photoproduction at $E_\gamma{\le}$6~GeV 
comes from the strong pion exchange contribution because the
ratio of $\omega\gamma\pi$ to $\rho\gamma\pi$ coupling constants
squared accounts for $\simeq$5.8. The solid lines in Fig.\ref{reg12}a)
and Fig.\ref{reg5a}e) show the calculation~\cite{Sibirtsev2} 
with meson exchange model, which as well includes $\pi$, $\eta$,
$\sigma$ and $N$ exchanges. Our calculations well reproduce the data
at $E_\gamma{\le}$5~GeV and indicate that this energy range 
can be entirely addressed by the standard meson exchange approach.
Apparently, low energy data can be well fitted by Regge model
with inclusion of $\pi$ and $f_2$ exchanges as is shown by the 
dashed line in Fig.\ref{reg5a}e). 

Finally, the data on $\omega$ photoproduction at high energies can be
well reproduced by Regge model when readjusting the Pomeron
exchange amplitude parameters. However within the same set 
of parameters and with experimental ${\gamma\omega}$ and 
${\gamma\rho}$ couplings it is not possible to describe 
simultaneously $\omega$ and $\rho$ photoproduction data.
Data on $\omega$ photoproduction at $E_\gamma{\le}$5~GeV
indicate a substantial contribution from meson exchange
reactions and can be well reproduced by calculations with
$\pi$, $\eta$, $\sigma$ and $N$ exchanges~\cite{Sibirtsev2}.


\begin{thebibliography} {99}
\bibitem{Battaglieri}
        M. Battaglieri et al., Phys. Rev. Lett. {\bf 87}, 172002 (2001).
\bibitem{Donnachie1}
        A. Donnachie and P.V. Landshoff, Phys. Lett. B {\bf 348}, 213 (1995);
        Nucl. Phys. B {\bf 311}, 509 (1988); Phys. Lett. B {\bf 470}, 
        243 (1999).
\bibitem{Breitweg}
        J. Breitweg et al., Eur. Phys. J. C {\bf 1}, 81 (1998).
\bibitem{Anciant}
        E. Anciant et al.,  Phys. Rev. Lett. {\bf 85}, 4682 (2000).
\bibitem{Laget1}
        J.M. Laget, Phys. Lett. B {\bf 489}, 313 (2000). 
\bibitem{Sibirtsev2}
        A. Sibirtsev, K. Tsushima and S. Krewald, nucl-th/0202083,
        in preparation. 
\bibitem{Renard}
        F.M. Renard, Nucl. Phys. B {\bf 15}, 267 (1970).
\bibitem{Weise}
        G. Gounaris and J.J. Sakurai, Phys. Rev. Lett. {\bf 21}, 244 (1968);
        F. Klingl, N. Kaiser and W. Weise, Z. Phys. A {\bf 356}, 193 (1996). 
\bibitem{Pichowsky}
        M.A. Pichowsky and T.S.H. Lee, Phys. Rev. D {\bf 56}, 1644 (1997).
\end{thebibliography}
\end{document}